\documentclass{article}
\usepackage{spconf,amsmath,graphicx}
\usepackage{amsfonts}
\usepackage[switch]{lineno}
\usepackage[named]{algo}
\usepackage[noend]{algpseudocode}
\usepackage{algorithm}
\usepackage{float}
\usepackage{hyperref}
\usepackage{multirow}
\usepackage{listings}
\usepackage{pifont}
\hypersetup{
    colorlinks=true,
    linkcolor=blue,
    filecolor=magenta,      
    urlcolor=blue,
    citecolor = blue,
}

\DeclareMathOperator*{\argmin}{\arg\min}
\DeclareMathOperator*{\argmax}{\arg\max}

\usepackage{xcolor}

\title{Single Snapshot Distillation for Phase Coded Mask Design in Phase Retrieval\\ \thanks{T\MakeLowercase{his work was supported by} Minciencias AMSUD 2023, CTO N\MakeLowercase{o. 112721-110-2024 under project} "C\MakeLowercase{omputational ultrasound imaging: from theory to applications", by the} VIE \MakeLowercase{from} UIS \MakeLowercase{under Project 3925 and by the }A\MakeLowercase{rmy} R\MakeLowercase{esearch Office/Laboratory under Grant Number }W911NF-25-1-0165, VIE \MakeLowercase{from} UIS \MakeLowercase{project 8087.} T\MakeLowercase{he views and conclusions contained in this document are those of the authors and should not be interpreted as representing the official policies, either expressed or implied, of the }A\MakeLowercase{rmy} R\MakeLowercase{esearch} L\MakeLowercase{aboratory or the} U.S. G\MakeLowercase{overnment.}}}

\name{Karen Fonseca$^{\star}$,\;Leon Suarez-Rodriguez$^{\dagger}$,\;Andrés Jerez$^{\dagger}$,\; Felipe Gutierrez-Barragan$^{\ddagger}$,\;Henry Arguello$^{\dagger}$}
\address{$^{\star}$ Department of Electrical Engineering, Universidad Industrial de Santander.\\
$^{\dagger}$ Department of Computer Science, Universidad Industrial de Santander.\\
$^{\ddagger}$ Independent Researcher.}

\begin{document}
\maketitle
\begin{abstract}
Phase retrieval (PR) reconstructs phase information from magnitude measurements, known as coded diffraction patterns (CDPs), whose quality
depends on the number of snapshots captured using coded phase masks. High-quality phase estimation requires multiple snapshots, which is not desired for efficient PR systems. End-to-end frameworks enable joint optimization of the optical system and the recovery neural network. However, their application is constrained by physical implementation limitations. Additionally, the framework is prone to gradient vanishing issues related to its global optimization process. This paper introduces a Knowledge Distillation (KD) optimization approach to address these limitations. KD transfers knowledge from a larger, lower-constrained network (teacher) to a smaller, more efficient, and implementable network (student). In this method, the teacher, a PR system trained with multiple snapshots, distills its knowledge into a single-snapshot PR system, the student. The loss functions compare the CPMs and the feature space of the recovery network. Simulations demonstrate that this approach improves reconstruction performance compared to a PR system trained without the teacher's guidance.
\end{abstract}
\begin{figure*}[tb]
    \centering
    \includegraphics[width=\linewidth]{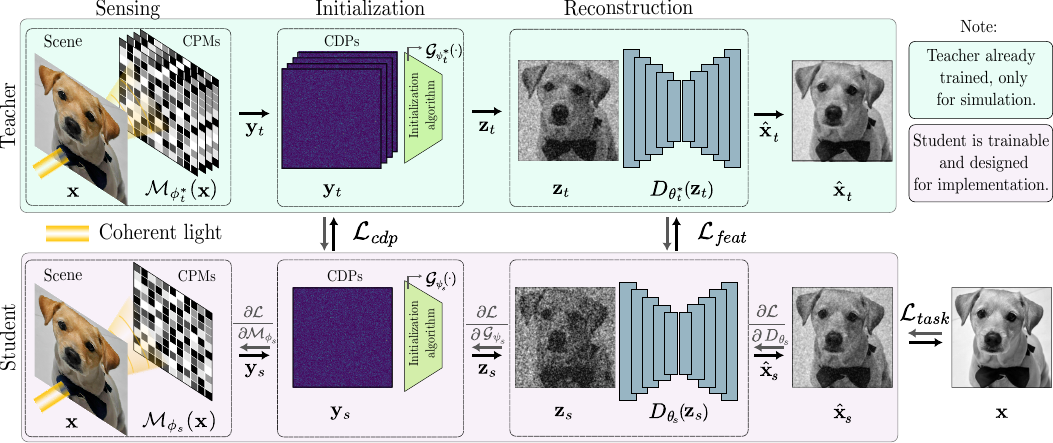}
    \caption{Method Description. The scene $\mathbf{x}$ is propagated through the optical system $\mathcal{M_\phi}(\cdot)$ illuminating with coherent light, producing the CDPs ($\mathbf{y}$). For simplicity, teacher and student representations are denoted by subscripts $t$ and $s$. An initialization process on the CDPs provides a first estimate of the scene, $\mathbf{z}$, which is then refined by the reconstruction network to obtain the final phase estimation, $\mathbf{\hat{x}}$. The teacher model is a trained, low-constrained, high-performance PR system, while the student is the single-snapshot PR system. During KD, two loss functions transfer knowledge from the teacher to the student: one for the CDPs and another for the reconstruction network’s bottleneck features.}
    \label{method}
\end{figure*}

\section{Introduction}
\label{sec:1_intro}
Phase retrieval (PR) is a nonlinear process that relates the intensity measurements to the underlying phase \cite{shechtman2015phase}. The ability to reconstruct the complex-valued optical field using only magnitude measurements is crucial in a wide range of scientific and engineering applications, including optics \cite{jerez2024deep}, astronomy \cite{astronomy}, and medical imaging \cite{MedicalImaging}. In optics, coded PR systems employ coded phase masks (CPMs) to modulate the phase, producing coded diffraction patterns (CDPs) captured by a detector. By acquiring multiple CDPs with different known phase modulations, the PR optimization problem becomes better conditioned \cite{schulz2021photon}. However, acquiring a large number of CDPs snapshots is impractical for efficient PR systems designed for dynamic scenarios. Key challenges in solving the PR problem include proper data initialization, CPMs design, and reconstruction algorithms. Furthermore, optical acquisition can influence the behavior of coded masks depending on the diffraction field \cite{pinilla2022unfolding}. Deep learning has become a powerful tool to address the PR problem \cite{wang2024use}. Recent state-of-the-art methods for CPM design leverage end-to-end (E2E) deep learning to jointly optimize CPMs and the PR reconstruction network, effectively reducing the required number of snapshots \cite{Ye:22}. To overcome this challenge, proper data initialization is critical to achieving accurate phase recovery. In \cite{jerez2024deep}, the authors propose learning the initialization process based on the E2E approach to approximate the complex-valued optical field.  However, in the E2E framework, CPMs implemented as the first layer of a deep neural network (DNN) can lead to vanishing gradients, resulting in suboptimal CPMs without proper regularization \cite{NEURIPS2019_851300ee}. Moreover, a single-snapshot system, which is highly constrained with fewer parameters than a multi-snapshot system, does not align with the empirical observation that overparameterized DNNs generally improve generalization \cite{allen2019learning}. To address these limitations, Knowledge Distillation (KD) \cite{hinton2015distillingknowledgeneuralnetwork} is explored. KD transfers knowledge from a well-conditioned, high-capacity teacher model to a lower-capacity student model. This process allows the smaller model to achieve comparable performance \cite{gou2021knowledge} by facilitating communication between models at different stages of the method, rather than relying solely on the network's outputs. KD has been successfully applied to high-level vision tasks such as classification \cite{hinton2015distillingknowledgeneuralnetwork}, object detection \cite{NIPS2017_e1e32e23}, and segmentation \cite{Liu_2019_CVPR}. KD has also been used in computational optical imaging to reduce the number of coded apertures in single-pixel imaging \cite{suarez2024highly}. In \cite{quan2023unsupervised}, snapshot reduction techniques have been explored to improve PR algorithms, including KD to reduce the size of neural networks. However, this approach only focuses on the recovery algorithm, without considering the optical design, which is crucial in optimizing the overall performance of PR systems. 

Therefore, we propose to improve the optimization of a single-snapshot PR system with KD-based criteria. The proposed method assumes that a learned, multi-snapshot E2E PR system (teacher) can guide the learning of a single-snapshot constrained PR system (student). This guidance is implemented through two KD loss functions, which compare the student's CDPs and feature space of the recovery network with those of the teacher networks during training, as illustrated in Figure \ref{method}. This training technique significantly improves the reconstruction accuracy of the learned single-snapshot system.

Our ablation study suggests that the two KD loss functions contribute to the improvement in reconstruction performance. However, applying KD in the first layer (CDP layer) has a greater impact on the performance than applying KD in the intermediate layers of the reconstruction network. This paper focuses on optimizing optical acquisition by reducing the number of snapshots through a multi-snapshot teacher model that trains a single-snapshot model, transferring the mathematical properties of well-conditioned systems to those with efficient acquisition.

The main contributions of this work are:
\begin{enumerate}
    \item Development of a KD optimization method designed for different phases of the E2E to enhance the performance of the learned single-snapshot phase recovery system. 
    \item Introduction of two knowledge transfer loss functions aimed at improving both the CPMs and the recovery network to ensure effective snapshot reduction through KD training.
    \item Comprehensive ablation study to identify the optimal student-teacher configuration for improved knowledge transfer and model efficiency.
\end{enumerate}
    
    
\section{End-to-End Phase retrieval system}
\subsection{Sensing model}
PR comprises a scene $\mathbf{x} \in \mathbb{C}^n$ that is acquired with a sensor that can only capture the magnitude $\mathbf{y}_{\ell} \in \mathbb{R}^n$, for $\ell \in \{1,.., L\}$ with $L$ being the total number of snapshots. This system consists of a coherent light that illuminates a scene over an optical field. Then, the optical field goes through the CPMs, and finally, the CDPs are acquired by a sensor located at a far diffraction field. This acquisition can be modeled as follows:
\begin{equation}
    \mathbf{y}_{\ell} = \vert \mathbf{F} \mathbf{A}_{\ell} \mathbf{x}\vert ^2 + \boldsymbol{\omega}_{\ell},
\end{equation}
\noindent where $\mathbf{A}_{\ell} \in \mathbb{C}^{n \times n}$ is the CPM for the $\ell$-th snapshot, given by $ \mathbf{A}_{\ell} = \mathrm{diag}{(e^{j\boldsymbol{\phi}_{\ell}})^\mathcal{H}}$,  where $\mathcal{H}$ denotes the transpose conjugate operator and $\boldsymbol{\phi}_{\ell}\in [0,2\pi]^n$ stands for the parameters in the sensing matrix. $\mathbf{F} \in \mathbb{C}^{n\times n}$  corresponds to the Fourier matrix, and $\boldsymbol{\omega}_{\ell}$ is the noise. Throughout this article, we define the operator $\mathcal{M}_\phi(\cdot)$ as the sensing operator where $\mathcal{M}_\phi(\mathbf{x})=[\vert \mathbf{F} \mathbf{A}_{1}\mathbf{x}\vert ^2, ...,\vert \mathbf{F} \mathbf{A}_{L}\mathbf{x}\vert ^2]$.

\subsection{Initialization step}
The initial scene estimation plays a crucial role in phase recovery. In this work, it is obtained through an initialization process based on a filter spectral initialization \cite{jerez2024deep}, which is represented by the following optimization problem:
\begin{equation}
    \mathbf{z}^* = \argmax_{\lVert \mathbf{z} \rVert^2 = 1} \mathbf{z}^\mathcal{H} \boldsymbol{\Gamma} \mathbf{z},
\end{equation}
\noindent where $\mathbf{z}^* \in \mathbb{C}^n$ is the optimal value of $\mathbf{z}$ which is the approximation of the original scene $\mathbf{x}$. The matrix $\boldsymbol{\Gamma}$ is given by calculating a low-pass version leading eigenvector of the measurements $\mathbf{y}$. This optimization problem is solved based on the power iteration method \cite{chen2015solving} as shown in algorithm \ref{init} as follows:

\begin{algorithm}[H]
\caption{Initialization}\label{alg:cap}
\begin{algorithmic}[1]
\small
    \State Input: Acquired data $\mathbf{y}$, maximum number of iterations $T$, and a low pass filter $\mathcal{G}_\psi(\cdot)$.
    
    \State $\mathbf{z} \leftarrow$ Chosen randomly.
    \State $\mathbf{\Gamma}(\mathbf{y})$  $\triangleright$ Compute.
    \For {$t=0: T-1$}
        \State $\mathbf{z}^{(t+1)}=\mathcal{G}_\psi\left(\boldsymbol{\Gamma} \mathbf{z}^{(t)}\right).\hspace{3mm} \triangleright$ Filtering step
    \State $\mathbf{z^*}^{(t+1)}=\frac{\mathbf{z}^{(t+1)}}{\left\|\mathbf{z}^{(t+1)}\right\|_2} . \hspace{2mm}\quad \triangleright$ Normalization step
    \EndFor    
    \State Return: Initial guess $\mathbf{z^*}$.
\end{algorithmic}
\label{init}
\end{algorithm}

\noindent where $\psi \in \mathbb{R}^{k\times k}$ with $k \in \mathbb{N}$ as the kernel size, stands for the parameters of the initialization filter $\mathcal{G}_\psi(\cdot)$.

\subsection{Optimization problem}

Traditional optimization of PR systems employs an E2E optimization approach, in which the sensing model parameters are incorporated into neural network reconstruction architectures as a trainable layer. The sensing model and the reconstruction network are then jointly trained to recover the phase information. As illustrated in Fig. \ref{method}, the phase mask operator $\mathcal{M}_{\phi}(\cdot)$ and the filter operator $\mathcal{G}_{\psi}(\cdot)$ include the trainable parameters $\phi$ and $\psi$, respectively, corresponding to the sensing model and the initialization step. The recovery neural network $\mathcal{D}_\theta$ includes the trainable parameters $\theta$. The recovery loss function is given by:
\begin{equation}
\mathcal{L}_{task}(\phi, \psi, \theta)= \frac{1}{Q}\sum_{q=1}^Q\|  
\mathcal{D}_\theta(\mathcal{G}_\psi( \mathcal{M}_\phi(\mathbf{x}_q)))  - \mathbf{x}_q \|_2^2,
\label{eq:task}
\end{equation}
\noindent where $Q$ is the total number of scenes, and  $q$ is the index of a specific scene. 

\noindent The PR E2E optimization problem is formulated as:
\begin{align}
    \{\phi, \psi, \theta\} = \; &\underset{\substack{\phi, \psi,\theta}} { \argmin} \; \alpha \mathcal{L}_{task}(\phi, \psi, \theta)  + \rho \mathcal{R}(\phi),
\label{eq.e2e}
\end{align}

\noindent where $\mathcal{R}$ is a regularization function that constrains the phase mask values. On the other hand, $\alpha$ and $\rho$ are the regularization parameters that weigh each term.
\section{Knowledge distillation for the design of a one-snapshot phase retrieval system}
To obtain high-quality phase estimation in PR requires multiple snapshots \cite{pinilla2022unfolding}. Furthermore, in traditional E2E optimization, the sensing model layer is prone to gradient vanishing problems, as it is the first layer of the E2E optimization framework \cite{NEURIPS2019_851300ee}. To address these limitations, we propose using a KD-based optimization approach to design a single-snapshot PR system. Specifically, we use as a teacher system a PR system with $L_t$ snapshots, to supervise the training of a single-snapshot student PR system through two knowledge transfer functions. 
\vspace{-0.3cm}
\subsection{Distillation Loss Functions}
To transfer the knowledge from the teacher to the student, we employ two loss functions, one focused on improving the design of the student's acquisition system parameters, and the other focused on improving the feature space of the student's recovery network, as illustrated in Fig. \ref{method}.
To enhance the design of the student's phase mask $\phi_s$ with $s$ snapshots, we propose the knowledge transfer function $\mathcal{L}_{cdp}$, which guides the average of the student CDPs measurements to approximate the average of the teacher's CDPs measurements. Furthermore, this loss function mitigates the vanishing gradient issue by introducing additional gradient pathways. The proposed loss function is given by:
\begin{equation}
    \mathcal{L}_{cdp}(\phi) = \frac{1}{Q}\sum_{q=1}^Q \| \frac{1}{L_t} \sum_{\ell_t=1}^{L_t} \mathbf{y}_{t[q,\ell_t]}-\frac{1}{L_s} \sum_{\ell_s=1}^{L_s}\mathbf{y}_{s[q,\ell_s]} \|_2^2,
\label{eq:cdp}
\end{equation}
\noindent where $\mathbf{y}_{t[q,\ell_t]}$ represents the teacher measurements of the scene $q$ in the snapshot number $\ell_t$, and $\mathbf{y}_{s[q,\ell_s]}$ represents the student measurements of the scene $q$ in the snapshot number $\ell_s$. To enhance the feature space of the student's reconstruction network, we leverage the intermediate features of the teacher model for guidance. The teacher's reconstruction network, trained with a sensing model that produces images with fewer gradations, extracts clearer features compared to the student's noisier features, which are constrained by the single-snapshot system. To address this, we align the bottleneck of the student's reconstruction network with that of the teacher, as the bottleneck provides the most compact feature space representation of the input image, given by:
\begin{equation}
    \mathcal{L}_{feat}(\phi,\psi,\theta)=  \frac{1}{Q}\sum_{q=1}^Q  \|\mathcal{D}_{\theta_t^*}(\mathbf{z}_{q,t})^{[b]} - \mathcal{D}_{\theta_s}(\mathbf{z}_{q,s})^{[b]}\|_2^2,
\label{eq:features}
\end{equation}
\noindent where $\mathcal{D}_{\theta_t^*}(\mathbf{z}_{q,t})^{[b]}$ and $\mathcal{D}_{\theta_s}(\mathbf{z}_{q,s})^{[b]}$ are the intermediate feature maps obtained from the teacher and student recovery network in the bottleneck layer $[b]$ in the scene $q$.

\subsection{Optimization problem for knowledge distillation in phase retrieval}

Given the defined loss functions denoted in Eq. \ref{eq:task}, \ref{eq:cdp}, and \ref{eq:features}, the KD optimization problem is defined as:

\begin{align}
    \{\phi, \psi&,\theta\}  =  \; \underset{\phi, \psi, \theta}{\arg\min} \;  \alpha \;\mathcal{L}_{task}(\phi, \psi, \theta)\;+ \notag \\  
    & \rho \mathcal{R}(\phi) + \beta \mathcal{L}_{cdp}(\phi) + \sigma \mathcal{L}_{feat}(\phi,\psi,\theta),\label{eq.KD}
\end{align}

\noindent where $\alpha$, $\beta$, and $\sigma$ are the regularization parameters that weight the loss terms, with $\sigma$ given by $\sigma = (1 - \alpha - \beta)$. These parameters are an affine combination.

\section{Simulations \& results}

\subsection{Training details:} In this work, we use the FashionMNIST dataset \cite{fashionmnist}, which contains images from 10 classes of clothing. The dataset consists of 60,000 images, split into 30,000 for training and 30,000 for validation, with an additional test set of 10,000 images. All images were resized to \(96 \times 96\). 
For the experiments, four different model configurations were trained. The teacher model and the E2E baseline were trained using only the E2E criterion, following equation \ref{eq.e2e}, with all parameters optimized. In the random setting, the loss function \ref{eq.e2e} was used, where the $\phi$ and $\psi$ parameters remained fixed. Finally, the student model followed the KD optimization problem in \ref{eq.KD}, where all parameters were trainable. All models were trained for 120 epochs using the Adam optimizer \cite{Adam} with a learning rate of $5 \times 10^{-4}$ and a batch size of 64 images. We ran the experiments using a NVIDIA GeForce RTX 4090 GPU and the PyTorch library. The initialization step of all four configurations used $T=25$ iterations. For the experiments, the well-known U-Net architecture \cite{Unet} was used as the reconstruction network $\mathcal{D}_{\theta}$, as the focus is on developing a new methodology for optimizing PR systems rather than designing new reconstruction neural network architectures. Nevertheless, the architecture can be modified without compromising the effectiveness of the proposed approach by adapting the knowledge transfer loss function $\mathcal{L}_{feat}$.

\begin{figure}[!bt]
    \centering
\includegraphics[width=0.9\linewidth]{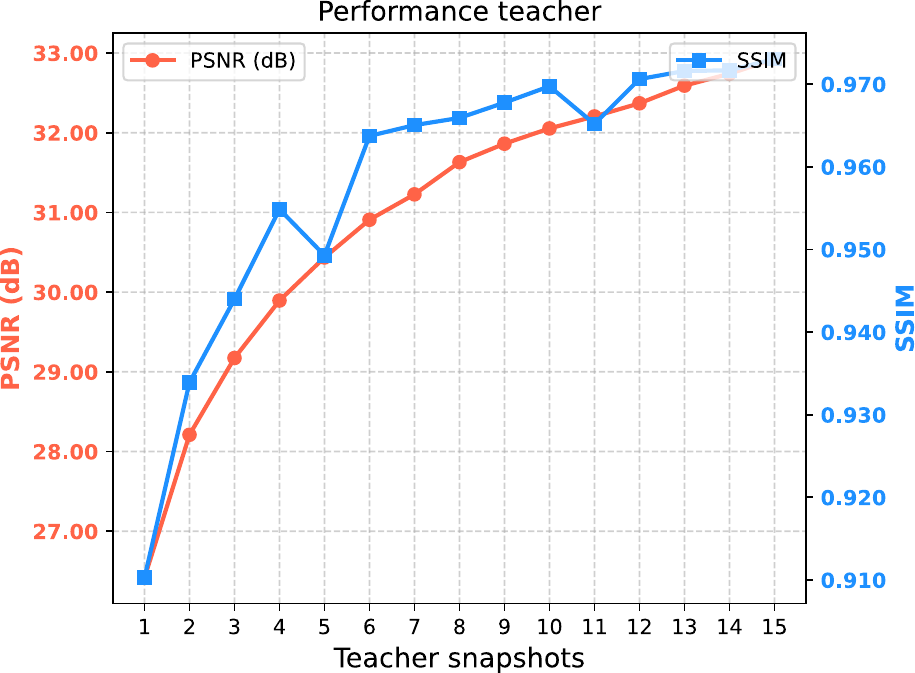}
    \caption{Teacher model's performance with different snapshots.}
    \label{fig:teacher-performance}
\end{figure}
\begin{figure}[!b]
    \centering
    \includegraphics[width=0.9\linewidth]{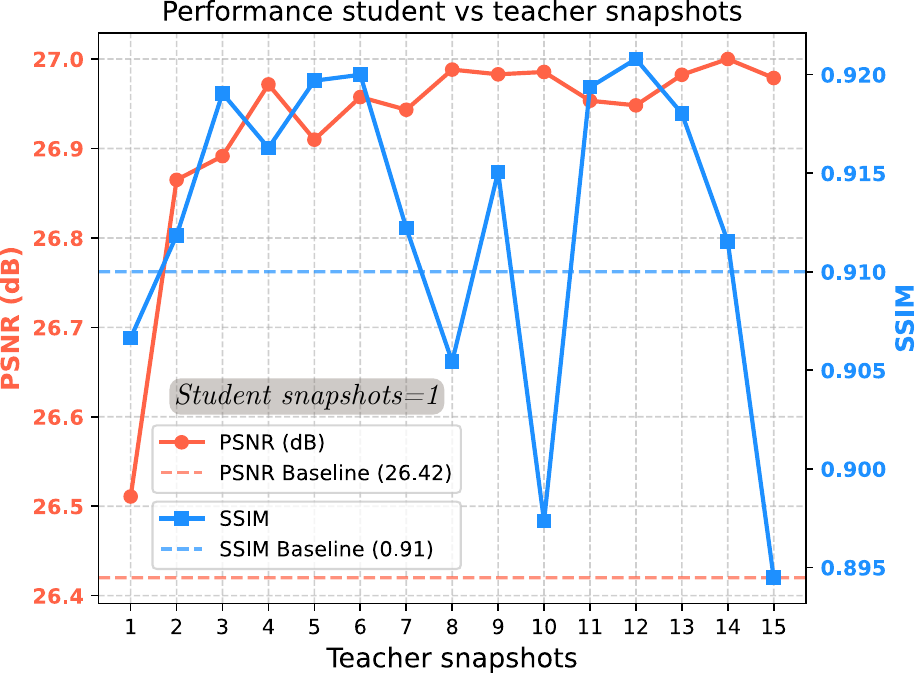}
    \caption{Student performance. The PSNR and SSIM from the student model for a single snapshot depend on the teacher model for multiple snapshots.}
    \label{fig:student-performance}
\end{figure}
\begin{figure*}[!t]
    \centering
    \includegraphics[width=\linewidth]{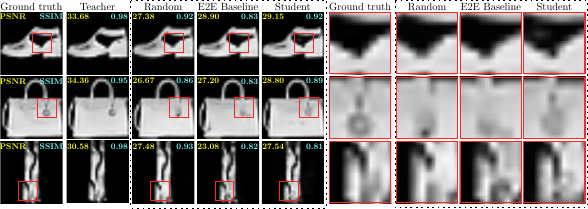}
    \caption{Reconstruction Results. Three reconstruction samples from the test set of the Fashion MNIST database are presented, illustrating the reconstructed images ($\mathbf{\hat{x}}$) along with their PSNR and SSIM values. The models included in the comparison are the teacher ($L_t = 9$), the random masks, the E2E baseline, and the student model (All before mention with $L_s = 1$).}
    \label{results_plot}
\end{figure*}
\subsection{One-snapshot student vs multiple snapshots teachers}
To determine the effect of the teacher setting on the student performance, several training sessions were conducted from 1 to 15 snapshots of the teacher. First, we train the teacher model and evaluate its performance at different snapshots to characterize its behavior over the training process (Fig. \ref{fig:teacher-performance}). Based on the previous experiment, and the evident improvement of the teacher with more snapshots, the KD was performed with different teachers, resulting in Fig.\ref{fig:student-performance}. Although the results demonstrate an improvement over the E2E baseline, increasing the number of teacher snapshots eventually leads to a performance plateau, with values stabilizing between 26.95 and 27 dB of PSNR. On the other hand, to verify the quality of the reconstructions, Fig. \ref{results_plot} was made to illustrate three different samples for a teacher of 15 snapshots (Fig. \ref{results_plot}). This experiment shows the reconstruction difference between the teacher ($L_t = 9$), the random, the E2E baseline, and student models $(L_s = 1)$. The results demonstrate an enhancement in PSNR metrics by up to 1.6 dB and 0.3 in SSIM, which is evident in the enhancement of image clipping details, as highlighted on the right side of the image.
\subsection{Ablation Studies}
In this section, we performed two ablation studies: one to determine the contribution of each independent knowledge transfer loss function to the overall performance gain and the other to assess the increase in computational resources during training. Table \ref{tab:ablation_kd_loss} presents the results on the contribution of each loss term, showing that the best PSNR results are obtained when both knowledge transfer loss functions, $\mathcal{L}_{cdp}$ and $\mathcal{L}_{feat}$, are combined, while the best SSIM results are achieved using only $\mathcal{L}_{cdp}$. These results were obtained with the teacher of $L_t = 9$ snapshots.
Furthermore, these results indicate that both loss functions contribute to performance improvement; however, the largest contribution comes from $\mathcal{L}_{cdp}$, as it directly enhances the design of the student's CPM, which determines the quality of the encoded information and its subsequent reconstruction.

\begin{table}[!tbh]
\centering
\small
\caption{Ablation study of the contribution of each independent term of the KD loss function}
\vspace{0.5cm}
\label{tab:ablation_kd_loss}
\begin{tabular}{ccccc}
\hline
\multicolumn{1}{c|}{$\mathcal{L}_{{cdp}}$} &\ding{51}&\ding{51}&\ding{56}&\ding{56}\\
\multicolumn{1}{c|}{$\mathcal{L}_{{feat}}$} &\ding{56}&\ding{51}&\ding{51}&\ding{56}\\
\multicolumn{1}{c|}{PSNR $\uparrow$}           & 26.94 & $\mathbf{26.98}$ & 26.70 & 26.41 \\
\multicolumn{1}{c|}{SSIM $\uparrow$}           & $\mathbf{0.923}$ & 0.915 & 0.912 & 0.910  \\ \hline
\end{tabular}
\end{table}

Table \ref{tab:training-time} compares computational resource usage during training between the proposed approach and E2E optimization. The results show a slight increase in training time of approximately one hour and an increase of about 1 GB in memory usage. However, despite this minor increase in computational resources during training, the performance gains justify it. Furthermore, we measured the inference time per batch to analyze the computational efficiency of the models. Our results indicate that the teacher model requires more than twice the inference time compared to the student model.

\begin{table}[!tbh]
\centering
\small
\caption{Comparison of the required training time and memory usage}
\begin{tabular}{clcc}
\cline{3-4}
\multicolumn{2}{c}{} & Teacher & Student \\ \hline
\multicolumn{2}{c}{Training Time $\downarrow$} & 6h 34m 18s & 7h 21m 25s \\
\multicolumn{2}{c}{Memory Usage (MB) $\downarrow$} & 5,577 & 6,647 \\
\multicolumn{2}{c}{Inference Time $\downarrow$} & 76 ms & 32 ms \\ \hline
\end{tabular}
\label{tab:training-time}
\end{table}
\vspace{-0.6
cm}
\section{Conclusion}
We propose an approach for optimizing PR systems inspired by the KD framework. By leveraging a more robust, high-performance multi-snapshot PR system as the teacher (which is inefficient for acquisition), we train a highly constrained, acquisition-efficient single-snapshot student PR system. Our findings show that extracting knowledge from both the teacher’s CPMs in the sensing model and its recovery network enables the student system to achieve improved reconstruction performance, surpassing traditional E2E optimization approaches for designing PR systems, even in the case of a single snapshot teacher. However, it is worth noting that increasing the teacher's size does not necessarily yield the best student configuration, as the intermediate number of snapshots produced better reconstruction metric values. Future work will explore the use of more complex datasets as well as other reconstruction algorithms and networks.

\clearpage

\bibliographystyle{IEEEbib}
\bibliography{refs}
\end{document}